\documentstyle[12pt,aaspp4]{article}

\newcommand{\hi}{H{\sc i}\ } 
\newcommand{\et}{PKS~1830$-$211\ }
\slugcomment{Accepted for publication in ApJ Letters, September 11 1996}

\lefthead{J.E.J.~Lovell {\em et.\ al.}}

\begin{document}

\title{PKS~1830$-$211: A Possible Compound Gravitational Lens}

\author{J.E.J.~Lovell\altaffilmark{1}, J.E.~Reynolds\altaffilmark{2},
  D.L.~Jauncey\altaffilmark{2}, P.R.~Backus\altaffilmark{3},
  P.M.~McCulloch\altaffilmark{1}, M.W.~Sinclair\altaffilmark{2},
  W.E.~Wilson\altaffilmark{2}, A.K.~Tzioumis\altaffilmark{2},
  E.A.~King\altaffilmark{2}, R.G.~Gough\altaffilmark{2},
  S.P.~Ellingsen\altaffilmark{1}, C.J.~Phillips\altaffilmark{1},
  R.A.~Preston\altaffilmark{4} \& D.L.~Jones\altaffilmark{4} \\}

\authoraddr{J.E.J. Lovell, Department of Physics, University of
  Tasmania, GPO Box 252C Hobart, Tasmania 7001, Australia}
\authoremail{jim.lovell@phys.utas.edu.au}

\altaffiltext{1}{Department of Physics, University of Tasmania,
  Hobart, Tasmania 7001, Australia} \altaffiltext{2}{Australia
  Telescope National Facility, CSIRO, Epping, New South Wales 2121,
  Australia} \altaffiltext{3}{SETI Institute, Mountain View,
  California, USA} \altaffiltext{4}{Jet Propulsion Laboratory,
  California Institute of Technology, Pasadena, California 91109, USA
  }

\begin{abstract}
  Measurements of the properties of gravitational lenses have the
  power to tell us what sort of universe we live in. The brightest
  known radio Einstein ring/gravitational lens PKS~1830$-$211
  (\cite{jau91}), whilst obscured by our Galaxy at optical
  wavelengths, has recently been shown to contain absorption at the
  millimetre waveband at a redshift of 0.89 (\cite{wik96}).  We report
  the detection of a new absorption feature, most likely due to
  neutral hydrogen in a second redshift system at $z = 0.19$.
  Follow-up VLBI observations have spatially resolved the absorption
  and reveal it to cover the NE compact component and part of the
  lower surface brightness ring.  This new information, together with
  existing evidence of the unusual VLBI radio structure and
  difficulties in modeling the lensing system, points to the
  existence of a second lensing galaxy along our line of sight and
  implies that PKS~1830$-$211 may be a compound gravitational lens.
\end{abstract}


\newpage

\section{Introduction}
The strong, $\sim$10~Jy, flat-spectrum radio source PKS~1830$-$211 was
first suggested to be a gravitational lens by Rao and Subrahmanyan
(1988). Three years later the source was identified as an Einstein
ring/gravitational lens (\cite{jau91}) and remains the brightest such
object found in the radio sky by almost two orders of magnitude.
While the interpretation of the source as a gravitational lens beyond
the Galaxy (\cite{sk92}) is secure, it lies in a crowded and heavily
obscured field close to the Galactic Center and so far all efforts to
identify optical or infra-red counterparts either for the lensing
galaxy or the lensed source have been unsuccessful (\cite{djo92};
\cite{jau93}). In particular, the failure of optical measurements to
furnish any redshifts has driven the search for these critical
parameters into the radio spectrum.

The symmetric morphology of the source, comprising two compact,
flat-spectrum components of similar brightness located on opposite
sides of a 1 arcsec ring, immediately suggests a close alignment of
the lensed source behind the lensing mass. Moreover, there is evidence
of unusually high rotation measures in some parts of the source which
argues that the lensing galaxy is probably a gas-rich spiral
(\cite{nnr93}), and suggests the possibility of detecting molecular
absorption.

\section{Observations and Interpretation}
Accordingly, we undertook a survey on 1995 June 10 and 11 for
redshifted \hi and OH absorption with the Parkes 64~m radio telescope,
as part of a cooperative observation programme with the Project
Phoenix group (\cite{tar96}). The Project Phoenix receiver and signal
processing equipment were used to cover a frequency range of 995 to
1675~MHz, nicely complementing a previous absorption search over the
frequency range 400$-$1000~MHz at Green Bank, which yielded a negative
result (\cite{mcm93}). Our observations excluded the two intervals
1535-1635~MHz and 1165-1175~MHz because of excessive interference.
Although interference of both terrestrial and satellite origin was
profuse over most of the remaining band, it was generally of very
narrow bandwidth and easily recognizable, and did not greatly impede
the search.

We detected only a single absorption feature with two sub-components
of similar amplitude, centered at 1191.1~MHz with an overall line width
of approximately 50~km/s (Figure 1a). The absorption feature was
detected with comparable strength on two consecutive days.  On the
second day, observations of PKS~1830$-$211 were bracketed with those
of PKS~1921$-$293, a nearby source of similar flux density. No
evidence of the absorption feature was seen in this comparison source.

VLBI observations were made on 1995 September 18 with four telescopes
of the Australian Long Baseline Array (LBA) (\cite{pre93};
\cite{jau94}); Hobart, Coonabarabran, Parkes and five antennas of the
Australia Telescope Compact Array (ATCA) acting as a phased array. S2
recorders (\cite{wei91}) were used, operating in dual polarization
(LCP and RCP) with 4~MHz bandwidth centered on 1191.0~MHz.  Correlated
visibilities between all six antennas of the ATCA were also recorded
to produce an improved total-power spectrum (Figure 1b).

PKS~1830$-$211 was observed over an 8 hour period interleaved with a
bandpass calibrator. The VLBI data were correlated at the ATNF VLBI
correlator (\cite{wil96}).  No correlated flux was detected on the
long ($>4\mbox{~M}\lambda$) baselines to Hobart owing to interstellar
scattering at this frequency (\cite{jon96}).

Our VLBI image in Figure 2 is similar to an earlier 2.3 GHz VLBI image
(\cite{jau91}) and shows the two compact components of PKS~1830$-$211,
but little of the low brightness ring which is heavily resolved at
this resolution. Figure 2 also shows the absorption spectrum of each
component, clearly demonstrating that the two velocity components of
the absorption system obscure different parts of the source. The low
velocity component is also heavily resolved and therefore must be
obscuring only the extended ring while the high velocity component is
resolved only partially and covers the NE component but is weak or
absent in the SW component. This, together with a comparison of the
relative optical depths in Figures 1 and 2, also allows us to infer
that the angular size of the absorbing features must be greater than a
few tenths of an arcsecond.

A molecular absorption system at a redshift of 0.88582 has already
been found in this source (\cite{wik96}), and is argued to arise in an
intervening galaxy rather than in the lensed object. This has been
confirmed by observations with the BIMA array which have spatially
resolved the $z = 0.89$ absorption (\cite{fry96}). We believe that our
detection constitutes a second absorption system in PKS~1830$-$211 and
is probably \hi absorption in an intervening galaxy at a redshift of
$0.1926 \pm 0.0001$. In support of this we note that there are no
catalogued lines near 2246~MHz, the rest frequency of our observed
absorption if it belongs to the $z = 0.89$ system. Further, we
observed the bright Galactic sources Sgr~B2, Ori~KL and IRC~10216 with
the 26~m telescope at Hobart to search for a possible unlisted
transition at this frequency, and detected nothing above an rms noise
level of $\sim$0.1\% of the continuum.  Moreover, our detection
appears to lie mainly in front of the NE component whilst the $z =
0.89$ absorption is confined to the SW component (\cite{wik96}; Frye
{\em et al.} 1996). Our absorption feature also displays velocity
structure not seen in the $z = 0.89$ absorption profiles.

The interpretation of the feature in Figure 1 as OH is unlikely for
two reasons. First, the spectra show none of the `satellite' profiles
typical of OH absorption and second, we see no evidence of absorption
at 1016~MHz corresponding to \hi at the same redshift, which we would
expect to be clearly visible. Neither can this be an hydrogen
recombination line as we would have detected many such lines across
the band and did not.

T. Wiklind and F. Combes (1996b) report no evidence of molecular
absorption in PKS~1830$-$211 at $z = 0.19$ in their SEST observations.
However this is not totally unexpected as the total solid angle
subtended by the source at these high frequencies is small
($\sim$1mas$^2$) and hence the probability of intersecting a dense
molecular cloud along the line of sight is presumably small.

The absorption feature seen in Figure 1 is very similar in both width
and column density to that found in the lens system 0218$+$357
(\cite{cry93}), which is convincingly argued to arise from \hi
absorption in the lensing galaxy, probably a spiral galaxy seen nearly
edge-on.  We favour a similar interpretation for the absorption in
PKS~1830$-$211, with the line of sight intercepting several \hi clouds
in a gas-rich galaxy at $z = 0.19$. The two features seen in the
absorption profile may well correspond to two spiral arms of $\sim$kpc
scale seen nearly superposed, both of which partially obscure the ring
and one of which obscures the NE compact component. Such a picture is
entirely consistent with the properties of \hi clouds and galaxy
dynamics observed within our own Galaxy.

Independent evidence for a considerable amount of material along the
line of sight is suggested by the unusually high rotation measure seen
in the NE component (Nair {\em et al.} 1993; \cite{sub90}).
Furthermore, the observed downturn in total flux density of
PKS~1830$-$211 below 1~GHz (\cite{rao88}) implies significant
free-free absorbing material obscuring the non-compact structure,
which has a steep spectrum and cannot be synchrotron self-absorbed.

\section{\et as a Compound Gravitational Lens}
While the \hi absorption toward PKS~1830$-$211 indicates the presence
of a galaxy at $z = 0.19$, this does not infallibly imply {\it a
  priori}\/ that gravitational lensing is taking place at this redshift,
any more than it does for the $z = 0.89$ system.  That lensing of some
kind is taking place is beyond dispute given, for example, the
near-simultaneous variation seen in the flux density variations of the
two compact components (\cite{van95}), apparently separated by more
than $\sim10$kpc: a most improbable effect in any non-lensed
interpretation. It seems almost certain therefore, that at least one
of the two redshift systems at $z = 0.19$ and $z = 0.89$ is partaking
in the lensing and there is further evidence that both systems may be
involved.

Firstly, the lens evinces a quite paradoxical appearance at high
resolution.  VLBI images made over a 3 year period (\cite{jon93};
\cite{gar96}) show the SW component to be unresolved at
$\sim$milliarcsecond resolution while the NE component shows a well
resolved linear structure. This difference in morphology of the two
components is too long-lived to be due to the difference in
propagation times to the two components, estimated to be no more than
a few tens of days (Nair {\em et al.} 1993; \cite{van95}). The results
presented here suggest that the second absorption system at $z = 0.19$
may be responsible for this striking disparity in the two images by
causing additional lensing distortion of the NE image, thus forming a
compound gravitational lens. The alternative explanation that
scattering by a high gas concentration in front of the NE component is
causing the additional linear structure is ruled out as there is no
$\lambda^2$ dependence on the length of this feature in the 5 and
15~GHz images (\cite{jon93} and \cite{gar96} respectively). Secondly,
attempts to model PKS~1830$-$211 with a single, simple gravitational
potential have achieved only modest success, and have produced
markedly different models. The models developed by Nair {\em et~al.}
(1993) and by Kochanek {\em et~al.} (1992), for example, differ in a
number of material respects, not least in the time delay between the
two compact components, which has the opposite sign in the Nair {\it
  et al.} model to that inferred from the Kochanek {\it et al.} model.

The presence of a second lensing galaxy in this system adds an extra
layer of complexity to any lensing model. This galaxy is likely to
influence the light travel time through the NE component and so must
be considered before ${\rm H}_{0}$ can be determined from the time
delay.  It is important therefore to estimate the mass and position of
the $z = 0.19$ system. We are unable to obtain this information from
our data. However polarization images of PKS~1830$-$211 at two or more
frequencies to map rotation measure would help delineate the
intervening material.

\acknowledgments The Australia Telescope is operated as a national
facility by CSIRO\@.  J.E.J.L and C.J.P. are supported by Australian
Postgraduate Research Awards. This research was carried out in part at
the Jet Propulsion Laboratory, California Institute of Technology,
under contract to the NASA.

\newpage

\begin{figure}[ht]
\plotfiddle{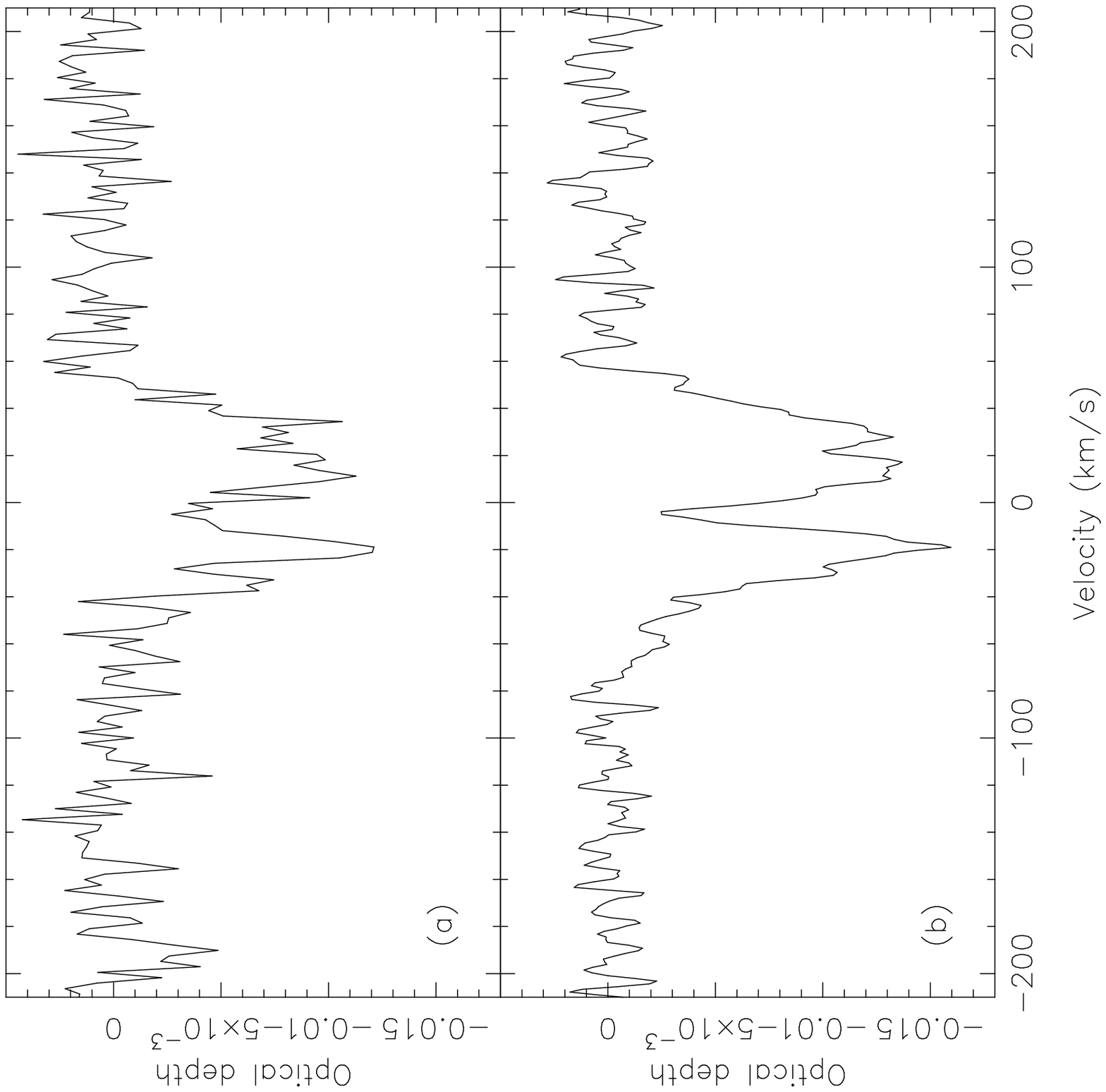}{0.8\textwidth}{270.0}{80.0}{80.0}{-226.0}{445}
\caption{a) Discovery spectrum taken at Parkes with the Project Phoenix
  SETI Receiver. b) Confirmation spectrum taken with the ATCA during our
  VLBI observations. The velocity scale on the horizontal axis assumes
  \hi as the molecular species.}
\end{figure}

\begin{figure}[ht]
\plotfiddle{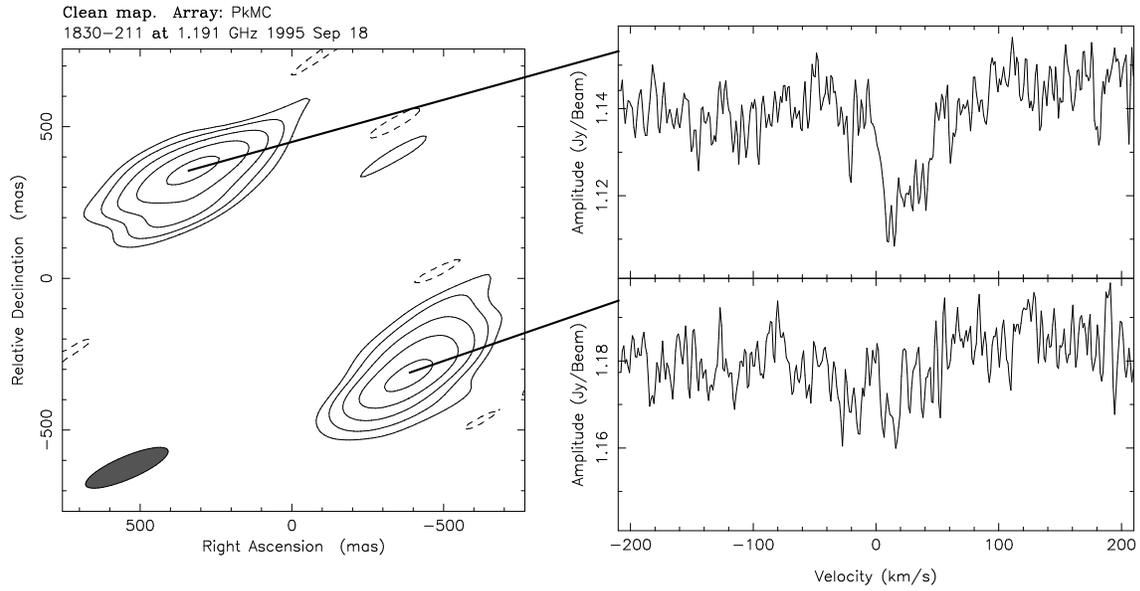}{0.1\textheight}{270.0}{60.0}{60.0}{-226.0}{260}
\caption{Our VLBI continuum map of PKS~1830$-$211 (left). The restoring beam
  is 294 by 75.2~mas and the contours are at $ -5, 5, 10, 20, 40$ and
  $80$ \% of the map peak which is 1.27~Jy per beam. Also shown here
  are the spectra at the positions of the two continuum components
  (right).}
\end{figure}

\end{document}